\documentclass[%
reprint,
amsmath,amssymb,
aps,
pra,
]{revtex4-1}

\usepackage{graphicx}
\usepackage{dcolumn}
\usepackage{bm}
\usepackage{amsmath}
\usepackage{enumerate}
\usepackage{setspace}
\usepackage{dsfont}
\usepackage{subfigure}
\usepackage{multirow}

\usepackage{threeparttable}
\usepackage[pdfstartview=FitH,
CJKbookmarks=true,
bookmarksnumbered=true,
bookmarksopen=true,
colorlinks, 
pdfborder=001,
linkcolor=blue,
anchorcolor=blue,
citecolor=blue
]{hyperref}

\begin{document}

	\title{Dual-frequency optical-microwave atomic clocks based on cesium atoms}
	
\author{Tiantian Shi$^{1,2,3}$, Qiang Wei$^{4}$, Xiaomin Qin$^{3}$, Zhenfeng Liu$^{3}$, Kunkun Chen$^{3}$, Shiying Cao$^{5}$, Hangbo Shi$^{3}$, Zijie Liu$^{3}$ \& Jingbiao Chen$^{3,6,*}$\footnote[*]{E-mail: jbchen@pku.edu.cn}}
	
	\address{$^1$School of Integrated Circuits, Peking University, Beijing 100871, China 
	\\
	$^2$National Key Laboratory of Advanced Micro and Nano Manufacture Technology, Beijing 100871, China 
	\\
	$^3$State Key Laboratory of Advanced Optical Communication Systems and Networks, Institute of Quantum Electronics, School of Electronics, Peking University, Beijing 100871, China 
	\\
	$^4$Chengdu Spaceon Electronics Co., Ltd., Chengdu 611731, China 
	\\
	$^5$National Institute of Metrology, Beijing 100029, China 
	\\
	$^6$National Laboratory for Quantum Information Science, Hefei 230026, China
	\\
	$^*$ jbchen@pku.edu.cn}

	\begin{abstract}
	
	\textbf{Abstract:} 
$^{133}$Cs, which is the only stable cesium (Cs) isotope, is one of the most investigated elements in atomic spectroscopy and was used to realize the atomic clock in 1955. Among all atomic clocks, the cesium atomic clock has a special place, since the current unit of time is based on a microwave transition in the Cs atom. In addition, the long lifetime of the $6{{\text{P}}_{{3 \mathord{\left/
				{\vphantom {3 2}} \right.
				\kern-\nulldelimiterspace} 2}}}$ state and simple preparation technique of Cs vapor cells have great relevance to quantum and atom optics experiments, which suggests the use of the $6{\text{S}} - 6{\text{P}}$ D2 transition as an optical frequency standard. In this work, using one laser as the local oscillator and Cs atoms as the quantum reference, we realized two atomic clocks in the optical and microwave frequencies, respectively. Both clocks could be freely switched or simultaneously output. The optical clock based on the vapor cell continuously operated with a frequency stability of $3.89 \times {10^{ - 13}}$ at 1 s, decreasing to $2.17 \times {10^{ - 13}}$ at 32 s, which was frequency stabilized by modulation transfer spectroscopy and estimated by an optical comb. Then, applying this stabilized laser for an optically pumped Cs beam atomic clock to reduce the laser frequency noise, we obtained a microwave clock with a frequency stability of ${{1.84 \times {{10}^{ - 12}}} \mathord{\left/
		{\vphantom {{1.84 \times {{10}^{ - 12}}} {\sqrt \tau  }}} \right.
		\kern-\nulldelimiterspace} {\sqrt \tau  }}$, reaching $5.99 \times {10^{ - 15}}$ at $10^5$ s. This study demonstrates an attractive feature for the commercialization and deployment of optical and microwave clocks and will guide further development of integrated atomic clocks with better stability. Thus, this study lays the groundwork for future quantum metrology and laser physics. 	
		
	\end{abstract}
	
	\date{\today}
	\maketitle

	\section*{Introduction}
	\label{Introduction}
	
The contributions of precise frequency measurements to global communications, satellite navigation and scientific research can hardly be overestimated, enabling us to build the most accurate atomic clocks\cite{Jun2022Resolving,Jun2019NP,Ludlow2018CM,PhysRevLett2019Al}. This feature allows atomic clocks to have a broad range of applications in basic physics research through frequency ratio measurements\cite{BACON2021Ratio}, such as the verification of general relativity\cite{Grotti2018Geodesy}, dark matter detection\cite{Roberts2020Search}, detection of changes in physical constants over time\cite{Barontini2022constants}, and definition and revision of a base unit of time in the International System of Units (SI)\cite{Riehle2018CIPM}. Although the measurement precision of the optical-lattice and single-ion clocks has reached the mid-19th digit\cite{Jun2019NP,Ludlow2018CM,PhysRevLett2019Al}, scientists continue to enrich the types of atomic clocks to probe for physics beyond the standard model, such as highly charged ions\cite{2018RevModPhysHCI}, single molecular ions\cite{Chou2017molecular}, vibrational molecular lattice\cite{2023PRXMolecular}, hydrogen optical lattice\cite{KawasakiPhysRevA2015}, mercury ion\cite{Ely2021Mecury}, Thorium-229 nuclear clock\cite{Kraemer2023Th229}, and even nature’s clock of millisecond radio pulsars \cite{Taylor1991pulsars}.


Despite the fact that there are many types of atomic clocks and that the Consultative Committee for Time and Frequency (CCTF) is committed to updating the roadmap toward the redefinition of the SI second\cite{Riehle2018CIPM,RIEHLE2015506,Dimarcq2024Metrologiae}, the current frequency standard realizations are based on the Cs atom. To maintain the continuity with the definition based on Cs, an ensemble definition can be created using the weighted geometrical mean of the frequencies of an ensemble of selected transitions, including Cs\cite{CIPM2022}. Therefore, the combination of Cs microwave clocks and optical clocks is widely applicable with their practical and performance advantages, respectively.

\begin{figure*}[t]
	\begin{center}
		\includegraphics[width=1\linewidth]{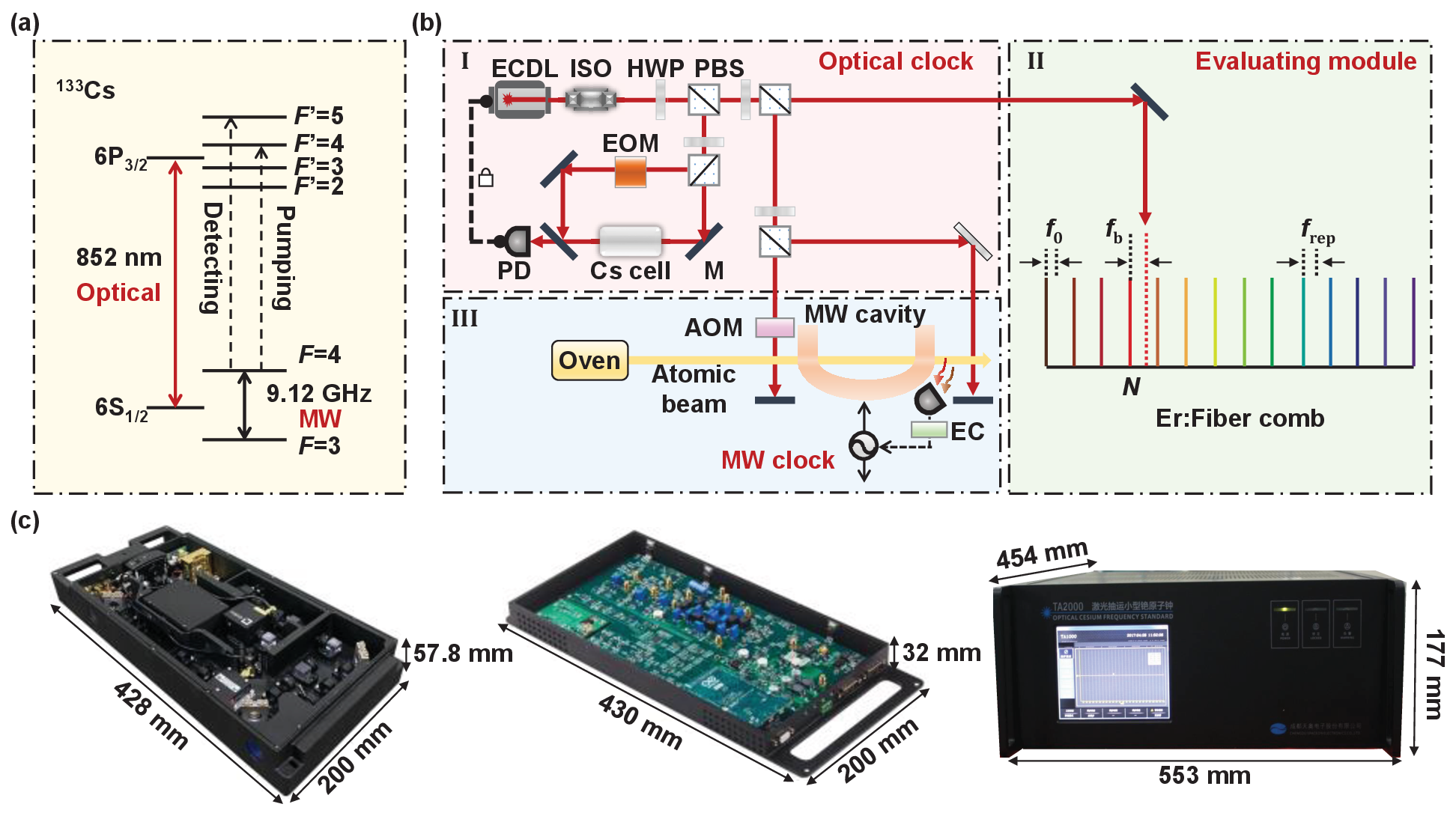}
	\end{center}
	\vspace{-5mm}
	\caption{{\bf Working principle of the dual-frequency optical-microwave atomic clocks based on cesium atoms.} {\bf a,} Level scheme for the $^{133}$Cs atom, which shows the 852-nm optical transition that was used to pump ($6{{\text{S}}_{{\text{1/2}}}}{\text{(}}F{\text{  =  4)}} - 6{{\text{P}}_{{\text{3/2}}}}{\text{(}}F'{\text{  =  4)}}$) and detect ($6{{\text{S}}_{{\text{1/2}}}}{\text{(}}F{\text{  =  4)}} - 6{{\text{P}}_{{\text{3/2}}}}{\text{(}}F'{\text{  =  5)}}$) for the Cs beam atomic clock. The 9.12-GHz microwave (MW) was used as the microwave clock signal. {\bf b,} Sketch of the optical-microwave atomic clock generation, which consists of three modules: I. The optical atomic clock generating module, II. optical atomic clock evaluating module, and III. microwave atomic clock generating module. An external-cavity diode laser (ECDL) was frequency-stabilized using the modulation transfer spectroscopy (MTS) technique. This laser source was divided into two parts. One beam beat with the ${N_{th}}$ tooth of an erbium-doped optical comb to estimate the frequency stability of the optical frequency standard with beat-note signal $f_b$, which was compared with a hydrogen maser. The initial frequency $f_0$ and repetition frequency ${f_{{\text{rep}}}}$ were locked to a hydrogen maser, whose frequency stability was $1 \times {10^{ - 13}}$ at 1 s. Then, the other beam was divided into two beams to pump and detect the lasers of the optically pumped Cs beam clock, respectively. Notations in the image: ISO: optical isolator; HWP: halfwave plate; PBS: polarized beam splitter; EOM: electro-optic modulation crystal; PD: photodetector; AOM: acousto-optic modulation crystal; MW: microwave; EC: electronic control module; M: high-reflectivity mirror. {\bf c,} Physical image of the dual-frequency optical-microwave atomic clocks. From left to right, there are the optics physical module, optics electrical module, and Cs microwave clocks. The optics physical and electrical modules were integrated into the microwave clock. The size of each module is indicated in the picture.}
	\label{Fig1}
\end{figure*}

Among all atomic clocks, the Cs clock is the most outstanding because it is used to define the SI second. Cs atoms have only one stable isotope $^{133}$Cs with a low moving speed at room temperature, which results in a narrow Doppler broadening. In addition, its hyperfine energy level is simple, so the Cs atomic clock has been used as the primary frequency standard since 1967. The long life time ($\tau  = 30.105(77)$ ns) \cite{1994PRAlifetime,1999PRALifetime,2003PRLlifetime} of the $6{{\text{P}}_{{3 \mathord{\left/
				{\vphantom {3 2}} \right.
				\kern-\nulldelimiterspace} 2}}}$ states is appealing because it provides a natural linewidth of approximately $2\pi  \times 5.234\left( {13} \right)$ MHz. In addition, Cs has many relevant physical and optical properties to various quantum optics experiments. First, optical clocks based on thermal atoms have a more compact structure and are easier to transport than optical lattice and ion clocks. Second, the frequency comparison among compact optical clocks is convenient and practical, which will enable significant advancements in various applications from laboratory to environment studies. Third, the Cs optical clock can be used as a high-performance laser source to integrate into an optically pumped cesium beam atomic clock to reduce the laser frequency noise and facilitate a high-performance microwave atomic clock. Thus, atomic clocks that use Cs atoms as the quantum references are of great significance.

Here, we used a Cs vapor cell and a Cs atomic beam as the frequency references to obtain atomic clocks in the optical and microwave and domains, respectively. Both clocks could be freely switched or simultaneously output. The physics package for the Cs optical clock was divided into optics and electronics boxes with dimensions of $428(W) \times 200(D) \times 57.8(H)$ mm$^3$ and $430(W) \times 200(D) \times 32(H)$ mm$^3$, respectively. A homemade external-cavity diode laser (ECDL) was frequency-stabilized to the Cs $6{{\text{S}}_{1/2}}(F = 4) - 6{{\text{P}}_{3/2}}(F' = 5)$ transition using modulation transfer spectroscopy (MTS) \cite{2011MTSNoh,2017MTSSchuldt,2020MTSShang,2022MTSHong,2022MTSmiao,2023MTSLee}. The frequency stability of the Cs optical clock was ${{{\text{4}} \times {{10}^{ - {\text{13}}}}} \mathord{\left/
		{\vphantom {{{\text{6}} \times {{10}^{ - {\text{13}}}}} {\sqrt \tau  }}} \right.
		\kern-\nulldelimiterspace} {\sqrt \tau  }}$, as measured by the erbium-doped optical comb, which was frequency referred to the active hydrogen maser. Furthermore, this compact optical frequency standard was used as the pump and probe lasers for the optically pumped Cs beam atomic clock, which had an overall size of $553(W) \times 454(D) \times 177(H)$ mm$^3$. The frequency stability of the Cs beam clock was ${{1.84 \times {{10}^{ - 12}}} \mathord{\left/
		{\vphantom {{1.7 \times {{10}^{ - 12}}} {\sqrt \tau  }}} \right.
		\kern-\nulldelimiterspace} {\sqrt \tau  }}$ and reached $5.99 \times {10^{ - 15}}$ at $10^5$ s. Since many physical and optical properties of Cs are relevant to various quantum optics experiments, this dual-frequency microwave-optical atomic clock based on thermal Cs atoms is technically ready for field applications. For example, it can serve as a timer for global satellite navigation and a high-performance light source for frequency comparison, which will have an immense impact on society.

\begin{figure*}[t]
\begin{center}\includegraphics[width=1\linewidth]{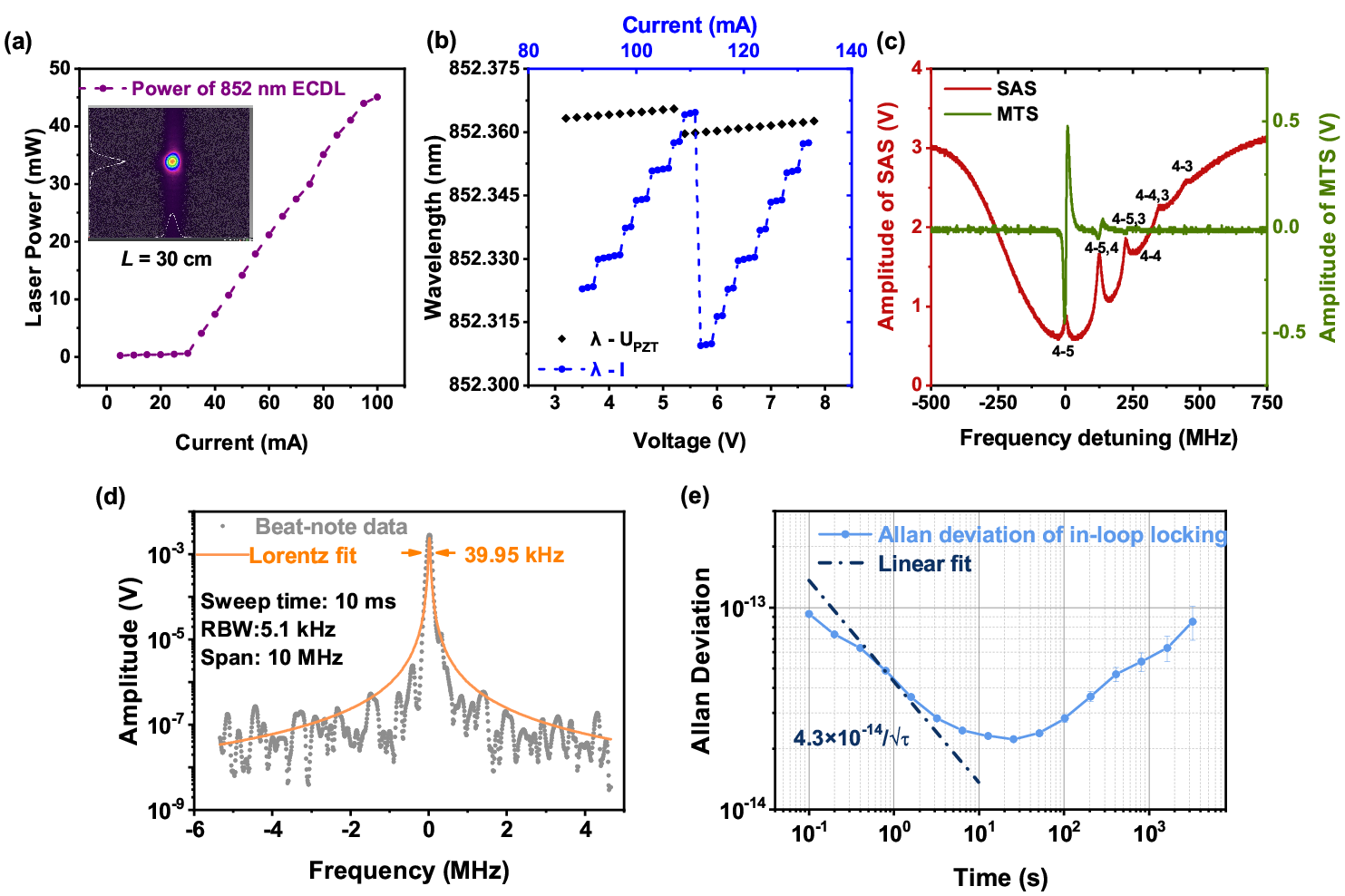}
	\end{center}
	\vspace{-5mm}
	\caption{{\bf Primary laser characteristics.} {\bf a,} Output laser power $P$ (purple dotted line) as a function of the current applied to the laser diode. The maximum lase power was 45 mW. Inset: light spot profile output from the ECDL approximately $L=30$ cm away from the external-cavity feedback mirror. It worked at the TEM00 mode for the next frequency stabilization. {\bf b,} Wavelengths of the ECDL with the change in injected current to the laser diode (blue dotted line) and voltage applied on the PZT (black dots). {\bf c,} Saturation absorption spectrum (red line) and the corresponding modulation transfer spectrum (green line) of the Cs $6{{\text{S}}_{{\text{1/2}}}}{\text{(}}F{\text{   =   4)}} - 6{{\text{P}}_{{\text{3/2}}}}{\text{(}}F'{\text{   =  3}},{\text{4}},{\text{5)}}$ transition. {\bf d,} Beat-note spectra (grey dots) between two identical 852-nm ECDLs. The resolution bandwidth was 5.1 kHz with a sweep time of 10 ms and a span of 10 MHz. Using Lorentz fitting, the fitted beat-note linewidth was $\left( {39.95 \pm 1.91} \right)$ kHz, which indicates that the linewidth of each ECDL was 28.25 kHz because the two ECDLs contributed equally to the laser noise. {\bf e,} Allan deviation of the MTS error signal after locking. It reflects the stability of the inloop locking, i.e., the tracking accuracy between ECDL and transition frequency of the reference atoms. The data were measured by recording the amplitude fluctuation of the error signal and transfer to the frequency fluctuation of the locking frequency point. The frequency stability of the in-loop locking (light blue line) was $9.28 \times {10^{ - 14}}$ at 0.1 s and decreased to $2.22 \times {10^{ - 14}}$ at 26 s. The dark blue dotted line represents the linear fitting of experimental data, and the result was ${4.3 \times {{10}^{ - 14}}}/{\sqrt \tau  }$.}
	\label{Fig2}
\end{figure*}

\section*{Results}
\label{Results}

{\bf Experimental scheme.} Here, we report an experimental demonstration of the dual-frequency optical-microwave atomic clocks. Fig. \ref{Fig1}a and Fig. \ref{Fig1}b show the energy level scheme and general setup, respectively. The transitions of Cs atoms at optical and microwave frequencies were used as reference standards of the optical and microwave clocks. For the optical clock, an ECDL working at a wavelength of 852 nm was used as the local oscillator, whose frequency was stabilized to the $6{{\text{S}}_{{\text{1/2}}}}{\text{(}}F{\text{  =  4)}} - 6{{\text{P}}_{{\text{3/2}}}}{\text{(}}F'{\text{  =  5)}}$ transition of thermal Cs atoms inside a vapor cell using the MTS technique, as shown in Fig. \ref{Fig1}b. Then, the frequency stability of the 852-nm optical frequency standard was estimated by optical heterodyne between the 852-nm laser and one tooth of the optical frequency comb, whose wavelength was nearly 852 nm. The initial frequency and repetition frequency of the optical comb were stabilized to a hydrogen maser. This stabilized low-noise 852-nm laser was also used as a light source of the optically pumped Cs beam clock. The 852-nm laser was divided into two parts: one part was frequency-shifted at approximately 251 MHz to the transition of Cs $6{{\text{S}}_{{\text{1/2}}}}{\text{(}}F{\text{  =  4)}} - 6{{\text{P}}_{{\text{3/2}}}}{\text{(}}F'{\text{  =  4)}}$ as the pumping laser; the other part was the detection laser of the microwave clock. By reducing the frequency noise of the laser source, we optimized the frequency stability of the Cs beam clock. In summary, we obtained a high-performance dual-wavelength microwave-optical atomic clock.

{\bf Cs optical frequency standard.} With the MTS technique\cite{Shirley:82}, i.e., transfer of modulation from the phase-modulated pump laser to the unmodulated probe laser, the modulated hole burning occurred in a sufficiently nonlinear resonant atomic medium in a vapor cell. Compared with saturation absorption spectroscopy (SAS), MTS reduces the direct feed-through of the laser amplitude noise while retaining most of the signal. In addition, MTS utilizes the advantage of high modulation frequencies and four-wave mixing to enhance the signal-to-noise performance above polarization spectroscopy. As an optically heterodyned saturation spectroscopy technique, MTS has the advantages of high resolution, high sensitivity, and being Doppler-free. Further, vapor-cell-based optical frequency standards feature a small size and low cost compared to optical frequency standards based on single ions and cold atoms, which promise a potential stability of $10^{-19}$ \cite{McGrew2018Nature,Oelker2019NP}, so vapor-cell-based optical frequency standards are expected to realize a transportable optical clock.  

Therefore, we demonstrate a simple optical clock frequency stabilized to the $6{{\text{S}}_{{\text{1/2}}}}{\text{(}}F{\text{  =  4)}} - 6{{\text{P}}_{{\text{3/2}}}}{\text{(}}F'{\text{  =  5)}}$ optical transition of Cs thermal atoms using the MTS technique. Cs atoms have a low melting point of only 28.5$^{\circ}$C, which can greatly reduce the power consumption. Moreover, the manufacture of a laser diode with an operation wavelength of 852 nm is mature. As shown in Fig. {\ref{Fig2}}(a), a homemade ECDL frequency selected by an interference filter was used as the local oscillator, whose laser power could reach 45 mW, which is sufficient for frequency stabilization and application. The laser worked in the TEM00 mode and was wavelength-tunable with the voltage of piezoelectric ceramic (PZT) to adjust the cavity length, and a current was applied on the laser diode, as depicted in Fig. {\ref{Fig2}}(b). To improve the environmental adaptability of the laser system, we designed the secondary temperature control for the laser diode with an accuracy of 0.01$^{\circ}$C. Next, the laser frequency was locked to the transition frequency of thermal Cs atoms. The double-layer Ni-Fe alloy and multi-layer Teflon were designed outside the vapor cell for magnetic shielding and heat retaining. Then, we locked the laser mode output from the 852-nm ECDL to the quantum reference using MTS. Through optimization, the power ratio between pumping laser and probe laser was 6:1, the vapor temperature was 35$^{\circ}$C to adapt to the high-temperature Cs oven inside a sealed atomic beam clock, the modulation frequency applied on the EOM was 4.97 MHz, and the slope of the MTS error signal was optimized to 0.07 V/MHz (Fig. {\ref{Fig2}}(c)). The full width at half maximum of the beating power spectroscopy was $39.95 \pm 1.91$ kHz as measured by the optical heterodyne between two identical stabilized ECDLs (Fig. {\ref{Fig2}}(d)). The in-loop frequency stability reached ${{4.3 \times {{10}^{ - 14}}} \mathord{\left/
		{\vphantom {{4.3 \times {{10}^{ - 14}}} {\sqrt \tau  }}} \right.
		\kern-\nulldelimiterspace} {\sqrt \tau  }}$ (Fig. {\ref{Fig2}}(e)). This was measured by recording fluctuations in the amplitude of the residual error signal after frequency locking, then converting the amplitude values into frequency values and calculating the in-loop frequency stability; this method shares many similarities with that proposed in \cite{ShiIEEE2022}. The in-loop locking accuracy can only reflect the tracking accuracy between local oscillator and frequency reference. The deterioration of frequency stability is mainly caused by mechanical vibration, temperature changes of the vapor cell and the EOM, and electrical locking noise. Moreover, we designed a function to automatically scan the atomic spectroscopy and relock again after losing lock to improve the reliability and increase the continuous locking time of the frequency-stabilized laser.

{\bf Frequency stability evaluation of the 852-nm laser using an optical frequency comb.} Here, we used an optical comb to estimate the frequency stability of the compact cell-based optical frequency standard. As shown in Fig. \ref{Fig1}(b), the frequency-stabilized 852-nm laser was beam-combined with the $N_{th}$ tooth of the erbium-doped optical comb to measure its frequency stability. The initial frequency $f_0$ and repetition frequency $f_r$ ($ \sim $200 MHz) were both locked to a hydrogen maser, so the phase-tracking stability and frequency instability of the optical comb were better than ${{5 \times {{10}^{ - 16}}} \mathord{\left/
		{\vphantom {{1 \times {{10}^{ - 15}}} {\sqrt \tau  }}} \right.
		\kern-\nulldelimiterspace} {\sqrt \tau  }}$ and ${{1 \times {{10}^{ - 13}}} \mathord{\left/
		{\vphantom {{1 \times {{10}^{ - 13}}} {\sqrt \tau  }}} \right.
		\kern-\nulldelimiterspace} {\sqrt \tau  }}$, respectively. Next, the frequency of the beating signal $f_b=f_N-f_{cw}$ was recorded by a frequency counter using the hydrogen maser as the frequency reference. The frequency stability of the hydrogen maser was better than ${{1 \times {{10}^{ - 13}}} \mathord{\left/
			{\vphantom {{1 \times {{10}^{ - 13}}} {\sqrt \tau  }}} \right.
			\kern-\nulldelimiterspace} {\sqrt \tau  }}$.

\begin{figure}[htbp]
	\centering\includegraphics[width=1\linewidth]{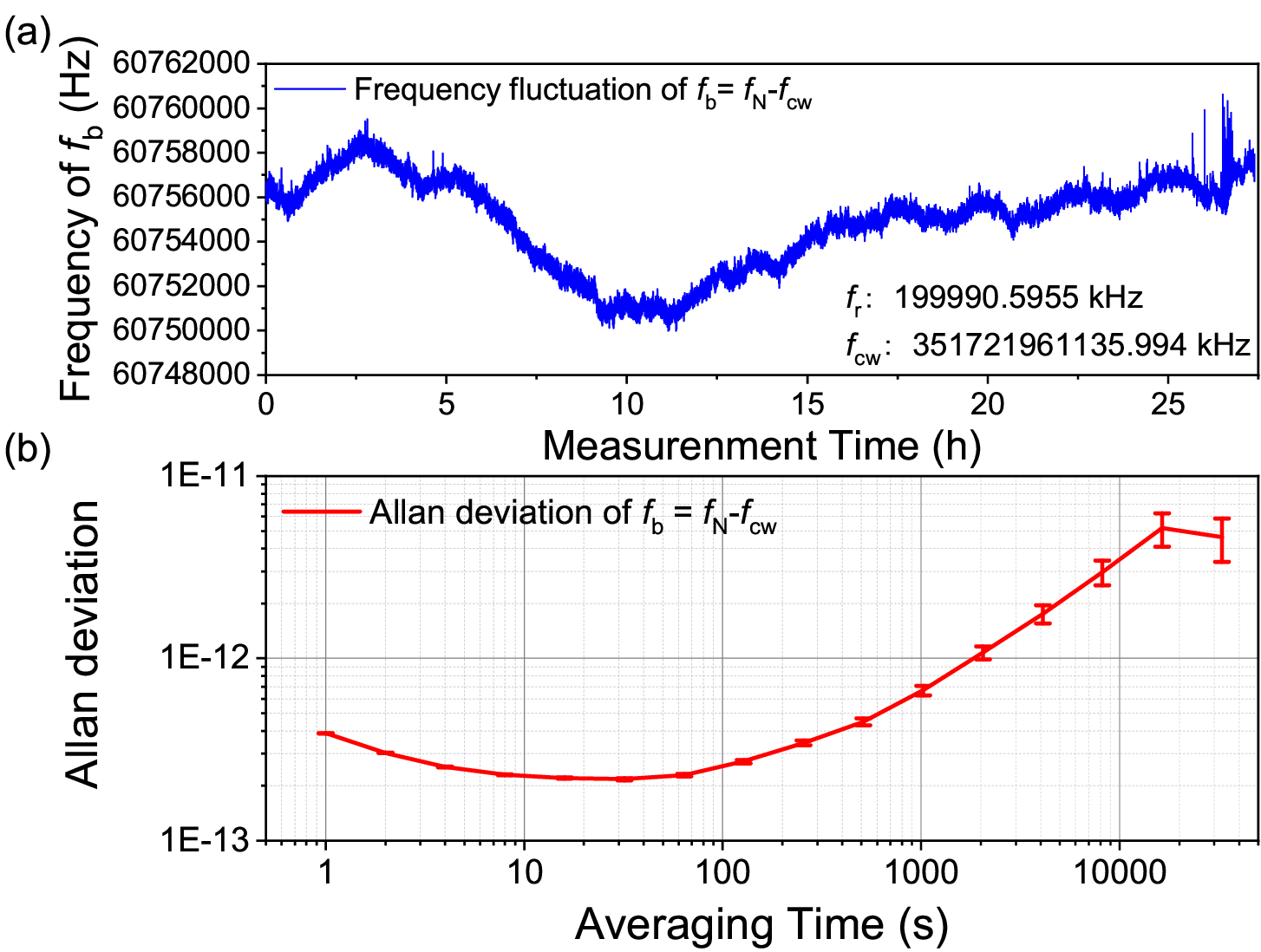}
	\caption{{\bf Frequency stability characteristic of the 852-nm optical frequency standard.} {\bf a,} Frequency fluctuation of the beat-note signal between the 852-nm optical laser and the $N_{th}$ laser mode of the optical comb frequency near the Cs $6{{\text{S}}_{{\text{1/2}}}}{\text{(}}F{\text{   =   4)}} - 6{{\text{P}}_{{\text{3/2}}}}{\text{(}}F'{\text{   =   5)}}$ transition. {\bf b,} Allan deviation of the beating frequency recorded in Fig. \ref{Fig4}(a). Frequency compared with a hydrogen maser; the frequency stability of the 852-nm optical frequency standard was $3.89 \times {10^{ - 13}}$ at 1 s and $2.17 \times {10^{ - 13}}$ at 32 s. After 100 s, the Allan deviation worsened because of the temperature drift of the atomic vapor temperature and power fluctuation of the pumping and probe lasers. Finally, the frequency stability was $5 \times {10^{ - 12}}$ over $10^4$ s.}
	\label{Fig3}
\end{figure}
		
		 After 28 hours of measurement, the frequency fluctuation of the stabilized laser was smaller than 8 kHz, as shown in Fig. \ref{Fig3}(a). Therefore, the Allan deviation of the 852-nm optical frequency standard was ${3.89 \times {{10}^{ - 13}}}$ at 1 s, ${2.17 \times {{10}^{ - 13}}}$ at 32 s, and ${3.3 \times {{10}^{ - 12}}}$ at $10^4$ s, as shown in Fig. \ref{Fig3}(b). The large frequency change within 10 hours resulted in a temperature drift of the atomic vapor cell that was used as the frequency reference. After a period of operation, the atomic temperature, i.e., the atomic number density, gradually stabilized, and the collision and Doppler shifts caused by atomic number fluctuation decreased. However, the frequency stability decreased to the best value of ${2.17 \times {{10}^{ - 13}}}$ at 32 s and worsened afterward because the long-term temperature drift of the reference atoms and electro-optic modulation crystal induced the frequency drift of the locking point and residual amplitude modulation (RAM) noise, respectively. Moreover, the power fluctuations of the pumping and probe lasers in MTS induced a light shift, which also worsened the long-term results. According to Refs.\cite{2017MTSSchuldt,2020MTSShang,2022MTSmiao,2023MTSLee}, the effects of the collision, Doppler shifts, light shift, and RAM noise are at the level of ${10^{ - 14}}$ in magnitude, which is consistent with the results of this work.

The optical signal stabilized by MTS is a frequency standard but not strictly an optical atomic clock. Nevertheless, this scheme reflects the performance of an optical signal and provides better frequency stability than most rf sources. In our next work, using a femtosecond-laser-based optical comb to provide the phase-coherent clock mechanism that links the optical and microwave frequencies, we will derive an rf clock signal of comparable stability over an extended wavelength range. The frequency stability of $4 \times {10^{ - 13}}$ at 1 s of a single optical oscillator at 852 nm will transfer to every comb tooth from 400 nm to 1600 nm. This work is of great significant to facilitate the optical frequency metrology using a compact and transportable clock in fundamental physics experiments and practical devices, such as optical communication.

{\bf Cs microwave atomic clock.} Using such a stable optical frequency standard as the pumping and repumping lasers for the Cs beam clock, the laser frequency noise was greatly reduced, and the performance of the Cs microwave atomic clock improved. (i) Compared with the distributed feedback Bragg (DFB) diode laser with SAS-stabilized frequency, the laser source obtained in this work has low frequency noise, and this reduces the noise to improve the signal-to-noise ratio (SNR) of the Ramsey signal. (ii) Although ECDL has lower frequency noise and a narrower laser linewidth \cite{Riehle2004Book}, it is more sensitive to mechanical vibration and easier to lose locking. Therefore, we designed the automatic lock electrical module. After locking has been lost, the auto-lock module will finely scan the voltage applied on the PZT that controls the external-cavity length or finely tune the control current applied on the laser diode to scan the MTS of the Cs $6{{\text{S}}_{{\text{1/2}}}}{\text{(}}F{\text{   =   4)}} - 6{{\text{P}}_{{\text{3/2}}}}{\text{(}}F'{\text{   =   5)}}$ transition and lock the laser frequency to this reference again. The relock time is less than 10 seconds. Therefore, the laser sources in this work simultaneously combines the advantages of low frequency noise and high reliability to optimize the frequency stability and continuous operation capability of Cs microwave clocks.  

\begin{figure}[htbp]
	\centering
	\includegraphics[width=0.95\linewidth]{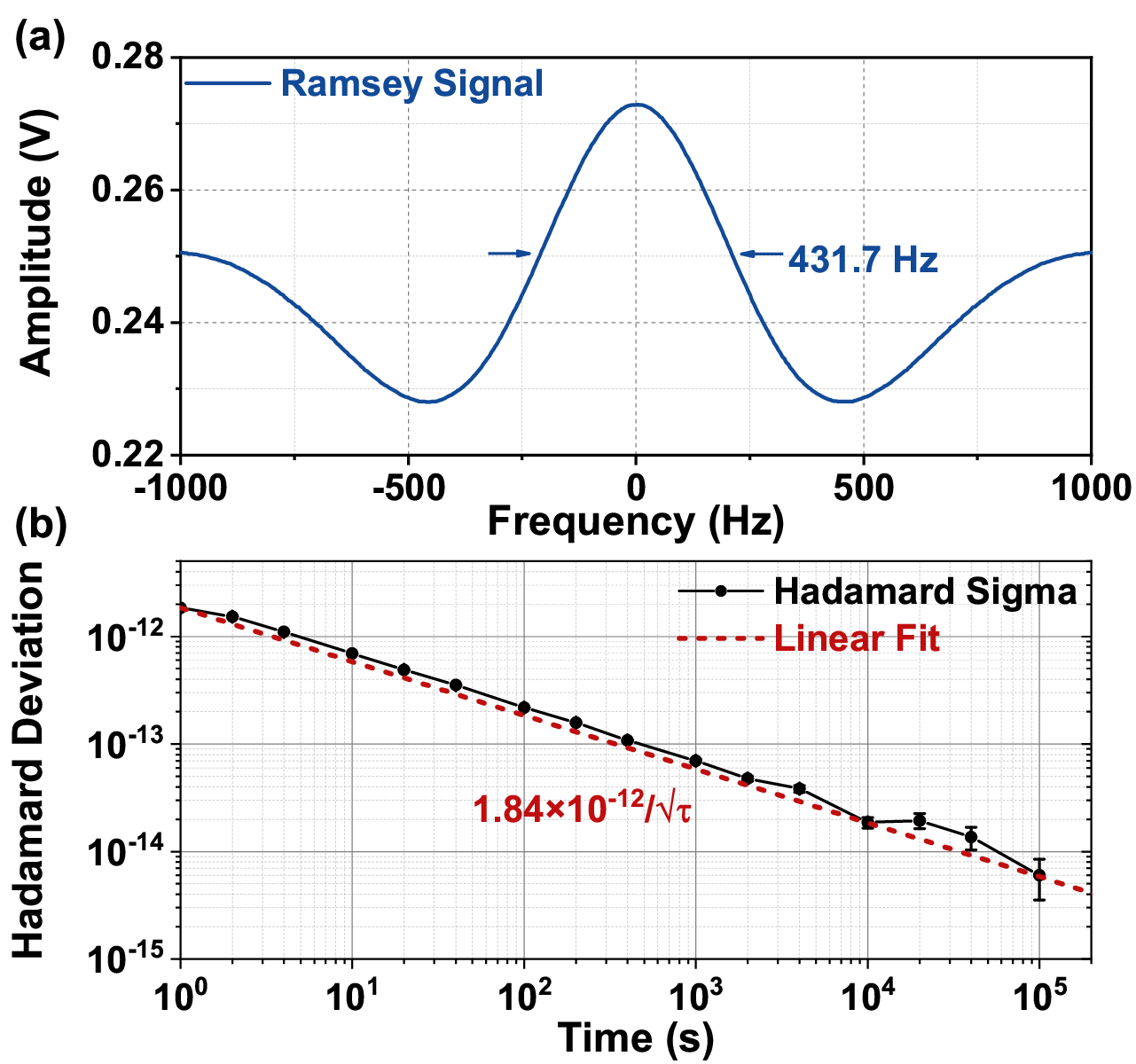}
	\caption{{\bf Frequency stability characteristic of a Cs microwave clock.} {\bf a,} Ramsey fringes of the Cs microwave clock transition (blue line). The full width at half maximum of the Ramsey central signal was 431.7 Hz. {\bf b,} Hadamard deviation of the Cs microwave clock. The 10-MHz frequency signal output from the optically pumped Cs beam clock was recorded by a frequency counter, which was referenced by the hydrogen maser. The Hadamard deviation (black dots) of the frequency was $1.84 \times {10^{ - 12}}$ at 1 s and $5.99 \times {10^{ - 15}}$ at $10^5$ s. It was linear fitted to ${1.84 \times {{10}^{ - 12}}}/{\sqrt \tau  }$, as shown by the red dotted line.}
	\label{Fig4}
\end{figure}

{\bf Frequency stability evaluation of the optically pumped Cs beam clock.} Figure \ref{Fig4}(a) shows the Ramsey fringes of the Cs microwave clock transition, whose linewidth is 431.7 Hz. The scanning range was 2 kHz, where the central frequency corresponded to the transition of ground state $\left| {F = 3,{m_F} = 0} \right\rangle  - \left| {F' = 4,{m_F} = 0} \right\rangle $. Using the hydrogen maser as a frequency reference, we measured the frequency stability of the compact Cs beam clock to be $1.84 \times {10^{ - 12}}$ at 1 s, $2.18 \times {10^{ - 13}}$ at 100 s, and $5.99 \times {10^{ - 15}}$ at $10^5$ s. These values are better than those of the magnetic selection state Cs beam clock and optically pumped Cs beam clock using a DFB laser\cite{XieCs2020} as the light source, as depicted in Fig. \ref{Fig4}(b). The frequency stability was calculated using Hadamard deviation because the frequency shift of the reference hydrogen clock was on the order of $5 \times {10^{ - 16}}$/day. During a testing time of more than 15 days, the magnitude of the reference clock's drift was similar to that of the long-term stability of the tested Cs microwave clock. Therefore, using Hadamard deviation for stability calculation is more reliable. Moreover, the temperature of the Cs oven was 100$^{\circ}$C, which is similar with other experiments. Therefore, the frequency stability was optimized due to the noise suppression of the laser sources. We integrated the frequency-stabilized 852-nm laser, including optical and electrical modules, into the Cs beam clock, and Fig. \ref{Fig1}(c) shows the Cs clock product with the size of $553(W) \times 454(D) \times 177(H)$ mm$^3$. This miniaturized atomic clock combines the advantages of excellent frequency stability and frequency accuracy, which is used for communication, traffic and national power as the primary frequency standard. Inside the sealed environment, the temperature of the 852-nm laser diode was seriously affected by the high-temperature oven, microwave generation module, and power supply, causing the internal temperature to be more than 30$^{\circ}$C above room temperature. The temperature fluctuation of the laser diode affected the laser wavelength, so we designed the double temperature control to improve the environmental adaptability of the laser. In the future, we will further optimize the immunity to mechanical vibration of the laser for a longer continuous locking time.

\section*{Conclusions}
\label{Conclusion}

In this work, we experimentally demonstrated a dual-frequency optical-microwave atomic clock based on thermal Cs atoms. Two simple and reliable atomic clocks were simultaneously built with convenient operation, which could arbitrarily switch or simultaneously output microwave and optical signals. In other words, they expanded the application range of practical microwave atomic clocks and miniaturized high-performance optical atomic clocks. First, using an ECDL as a local oscillator, after optimizing the experimental parameters, we achieved a highly stable Cs 852-nm optical frequency standard, with frequency stability being better than $7 \times {10^{ - 13}}$ between 1 s and 1000 s. The performance of this frequency standard was good compared to ${{\text{I}}_{\text{2}}}$ (short-term frequency stability) and Rb two-photon optical frequency standards, which is expected to become a portable compact optical atomic clock for gravitational wave detection, geodesy and measurement of time variation of fundamental constants. Second, by applying this optical frequency standard as the laser sources in the Cs microwave atomic clock, we optimized the frequency stability of the Cs microwave clock to be $5.99 \times {10^{ - 15}}$ at $10^5$ s, which is better than the values of most miniaturized microwave clocks. 

In the future, we will derive an rf clock signal of comparable frequency stability as the 852-nm optical frequency standard over an extended wavelength from 400 nm to 1600 nm. This result will make Cs atoms the next length standard, include them in the CIPM and facilitate the application of frequency metrology to precision experiments for fundamental physics and practical devices for communication.

\section*{Data availability}
All data that support the plots in this paper are available from the corresponding author upon request.

\bibliographystyle{elsarticle-num}
\bibliography{DFMicroOptAtomicClock}

	\section*{Acknowledgments}
	
	This study was funded by the National Natural Science Foundation of China (NSFC) (91436210), China Postdoctoral Science Foundation (BX2021020), Wenzhou Major Science \& Technology Innovation Key Project (ZG2020046), and Innovation Program for Quantum Science and Technology (2021ZD0303200).

	\section*{Author contributions}
	
	J.C. and T. S. conceived the idea to use thermal Cs atoms as the frequency reference to make two atomic clocks operate in both optical and microwave range. T.S., Z. L., Q. W., K. C., and S. C. performed the experiments. T.S. wrote the manuscript. J.C. and X. Q. provided revisions.

	\section*{Competing interests}
	
	The authors declare no competing interests.

	\section*{Additional information}
	
	
	
	{\bf Correspondence and requests for materials} should be addressed to T.S. or J.C.
	

	

\end{document}